\documentclass[sigconf]{acmart}
\AtBeginDocument{%
  }
    
\setcopyright{none}
\renewcommand\footnotetextcopyrightpermission[1]{}
\acmDOI{}
\acmISBN{}

\acmConference{}{}{}

\usepackage{enumitem}
\usepackage[ruled,vlined,linesnumbered]{algorithm2e}
\usepackage{multirow}
\usepackage{caption}
\usepackage{placeins}

\usepackage[english]{babel}
\usepackage{microtype}
\usepackage{subcaption}  
\renewcommand{\arraystretch}{0.85}
\author{Wuhan Chen}
\affiliation{
  \institution{School of Big Data \& Software Engineering, Chongqing University}
  \city{Chongqing}
  \country{China}
}
\email{wh.chen@cqu.edu.cn}

\author{Min Gao}
\authornote{Corresponding author.}
\affiliation{
  \institution{School of Big Data \& Software Engineering, Chongqing University}
  \city{Chongqing}
  \country{China}
}
\email{gaomin@cqu.edu.cn}

\author{Xin Xia}
\affiliation{
  \institution{School of Electrical Engineering and Computer Science, The University of Queensland}
  \city{brisbane}
  \country{Australia}
}
\email{x.xia@uq.edu.au}

\author{Zongwei Wang}
\affiliation{
  \institution{School of Big Data \& Software Engineering, Chongqing University}
  \city{Chongqing}
  \country{China}
}
\email{zongwei@cqu.edu.cn}

\author{Wentao Li}
\affiliation{
  \institution{University of Leicester}
  \country{United Kingdom}
}
\email{wl226@leicester.ac.uk}

\author{Shane Culpepper}
\affiliation{
  \institution{School of Electrical Engineering and Computer Science, The University of Queensland}
  \city{brisbane}
  \country{Australia}
}
\email{s.culpepper@uq.edu.au}
\begin{document}

\title[Multi-LLM Token Filtering and Routing for Sequential Recommendation]{Multi-LLM Token Filtering and Routing for Sequential Recommendation}

\renewcommand{\shortauthors}{}

\begin{abstract}
Large language models (LLMs) have recently shown promise in recommendation by providing rich semantic knowledge.
While most existing approaches rely on external textual corpora to align LLMs with recommender systems, we revisit a more fundamental yet underexplored question: \textbf{Can recommendation benefit from LLM token embeddings alone without textual input?}
Through a systematic empirical study, we show that directly injecting token embeddings from a single LLM into sequential recommenders leads to unstable or limited gains, due to semantic misalignment, insufficient task adaptation, and the restricted coverage of individual LLMs.
To address these challenges, we propose \textbf{MLTFR}, a \underline{M}ulti-\underline{L}LM \underline{T}oken \underline{F}iltering and \underline{R}outing framework for \textbf{corpus-free} sequential recommendation. MLTFR follows an interaction-guided LLM knowledge integration paradigm, where task-relevant token embeddings are selected via user-guided token filtering to suppress noisy and irrelevant vocabulary signals. To overcome the limitations of single-LLM representations, MLTFR integrates multiple LLM token spaces through a Mixture-of-Experts architecture, with a Fisher-weighted semantic consensus expert to balance heterogeneous experts and prevent domination during training.
By jointly filtering informative tokens and aggregating complementary semantic knowledge across multiple LLMs, MLTFR enables stable and effective utilization of LLM token embeddings without textual inputs or backbone modification. Extensive experiments demonstrate that MLTFR consistently outperforms state-of-the-art sequential recommendation baselines and existing alignment methods. Our code is available at: https://github.com/ccwwhhh/MLTFR.
\end{abstract}




\maketitle
\fancyhead[RO,RE]{}
\section{Introduction}
Recommender systems aim to learn representations of users and items from historical interactions to infer user interests and predict future behaviors, and have been widely deployed in real-world platforms such as e-commerce and media services~\cite{wang2023efficient, li2025linking, wang2019minimax}. Sequential recommendation, in particular, models the temporal dynamics of user behaviors and has achieved strong performance with Transformer-based architectures. 
Recently, the rapid progress of large language models (LLMs)~\cite{naveed2023comprehensive} has extended their influence beyond natural language processing. To leverage the rich semantic knowledge of LLMs for recommendation, a growing body of work integrates LLMs into downstream recommender systems, where LLMs serve as standalone recommenders or as semantic encoders that enrich item representations using textual attributes such as titles, descriptions, or reviews~\cite{zhai2025learning, hu2024enhancing, wang2025id}. While effective, these approaches rely heavily on the availability of high-quality textual corpora, which may be unavailable, incomplete, or costly to maintain in many practical settings.

\begin{figure}[t]
    \centering
  \includegraphics[width=1\linewidth]{figs/introduction_2.9.png}
    \caption{Illustration and empirical evidence of relevant token selection.
    The upper panel illustrates filtering noisy LLM token embeddings into task-relevant tokens. The lower panel compares different token embeddings on the SASRec backbone across two datasets. The dashed line represents the baseline performance without LLM integration. TE and RE denote token embeddings directly extracted from LLMs and randomly initialized embeddings, respectively.}
    \label{fig:tsc_intro}
    \vspace{-20pt}
\end{figure}

Beyond external text, LLMs inherently encode rich semantic knowledge within their parameters, particularly in their token embedding spaces, which capture entity-level semantics learned during large-scale pretraining~\cite{ju2024large, zhuang2024embedllm}. This observation suggests that LLM token embeddings may provide transferable semantic structure that complements ID-based item representations at low cost. Intuitively, since a user’s interaction history can be viewed as a sequence of item entities, token embeddings offer a natural medium for injecting semantic signals into sequential recommendation models, even in the absence of textual corpora. This raises a natural question: \emph{Can LLM token embeddings alone improve sequential recommendation performance?}

To investigate whether LLM token embeddings alone can improve sequential recommendation~\cite{sun2019bert4rec,kang2018self,li2025linking}, we study a \textbf{corpus-free} setting, where no item-level textual information is available during training or inference. As a preliminary attempt, we directly integrate token embeddings extracted from pretrained LLMs into a sequential recommender by projecting them to match the input embedding dimensionality and adding them to sequence embeddings via element-wise addition (denoted as \textit{LLM-TE}). As a control, we replace these token embeddings with randomly initialized embeddings of the same shape (\textit{LLM-RE}). 
As shown in Fig.~\ref{fig:tsc_intro}, directly injecting LLM token embeddings yields better performance than random embeddings, indicating that LLM token embeddings indeed encode transferable semantic information. However, the resulting gains are limited and unstable, and consistently underperform the original backbone. This observation suggests that na\"ively incorporating LLM token embeddings introduces substantial noise and semantic misalignment, motivating the need for task-relevant token selection.

We identify two fundamental challenges underlying this phenomenon. First, LLM vocabularies are large and generic, whereas user interests in sequential recommendation are sparse and task-specific; consequently, only a small fraction of token embeddings is relevant to a given user’s interaction history, while the majority behaves as noise. Second, relying on a single LLM provides a limited semantic perspective, leading to coverage gaps and unstable training dynamics when token embeddings are directly injected.

Motivated by these observations, we propose \underline{M}ulti-\underline{L}LM \underline{T}oken \underline{F}iltering and \underline{R}outing (\textbf{MLTFR}), a plug-and-play framework for corpus-free sequential recommendation. MLTFR follows an interaction-guided LLM knowledge integration paradigm, treating LLM token embeddings as lightweight semantic resources rather than full reasoning modules. To address the first challenge, MLTFR introduces a user-guided token filtering mechanism that selects a compact set of task-relevant tokens by measuring their relevance to user interaction representations, thereby suppressing noisy and irrelevant vocabulary signals. To address the second challenge, MLTFR integrates token embeddings from multiple LLMs through a Mixture-of-Experts (MoE) architecture~\cite{zhang2022mixture, ma2018modeling, jacobs1991adaptive}, enabling complementary semantic knowledge to be leveraged across models.
A key difficulty in multi-LLM integration is expert domination, where a single LLM disproportionately influences training and degrades robustness. To mitigate this issue, MLTFR introduces a semantic consensus expert based on Fisher-weighted aggregation~\cite{ly2017tutorial,gu2025delta}, which fuses shared token-level signals across LLMs while down-weighting conflicting cues~\cite{liu2024deepseek}. Combined with user-guided token filtering, this dual-module design enables stable and effective utilization of LLM token embeddings without requiring textual corpora or modifying backbone architectures.
Our contributions can be summarized as:
\begin{itemize}[leftmargin=*]
    \item We propose a \textbf{corpus-free} paradigm, \emph{Interaction-Guided LLM Knowledge Integration}, which leverages LLM token embeddings as lightweight semantic resources for sequential recommendation without relying on textual inputs.
    
    \item We introduce a \textbf{user-guided token filtering} strategy that selects task-relevant LLM tokens aligned with user interests, effectively suppressing noisy vocabulary signals and improving recommendation relevance.
    
    \item We develop a \textbf{shared expert MoE routing} mechanism to integrate multiple LLMs, forming a semantic consensus that mitigates cross-LLM conflicts and enables robust and adaptable performance.
\end{itemize}




\section{Methodology}
We study the standard next-item sequential recommendation task: given a user’s chronological interaction sequence $X_u = \{i_1^u, i_2^u, \ldots, i_k^u\}$, our goal is to predict the next item $i_{k+1}^u$.

\begin{figure*}
 \centering
\includegraphics[width=0.9\textwidth]{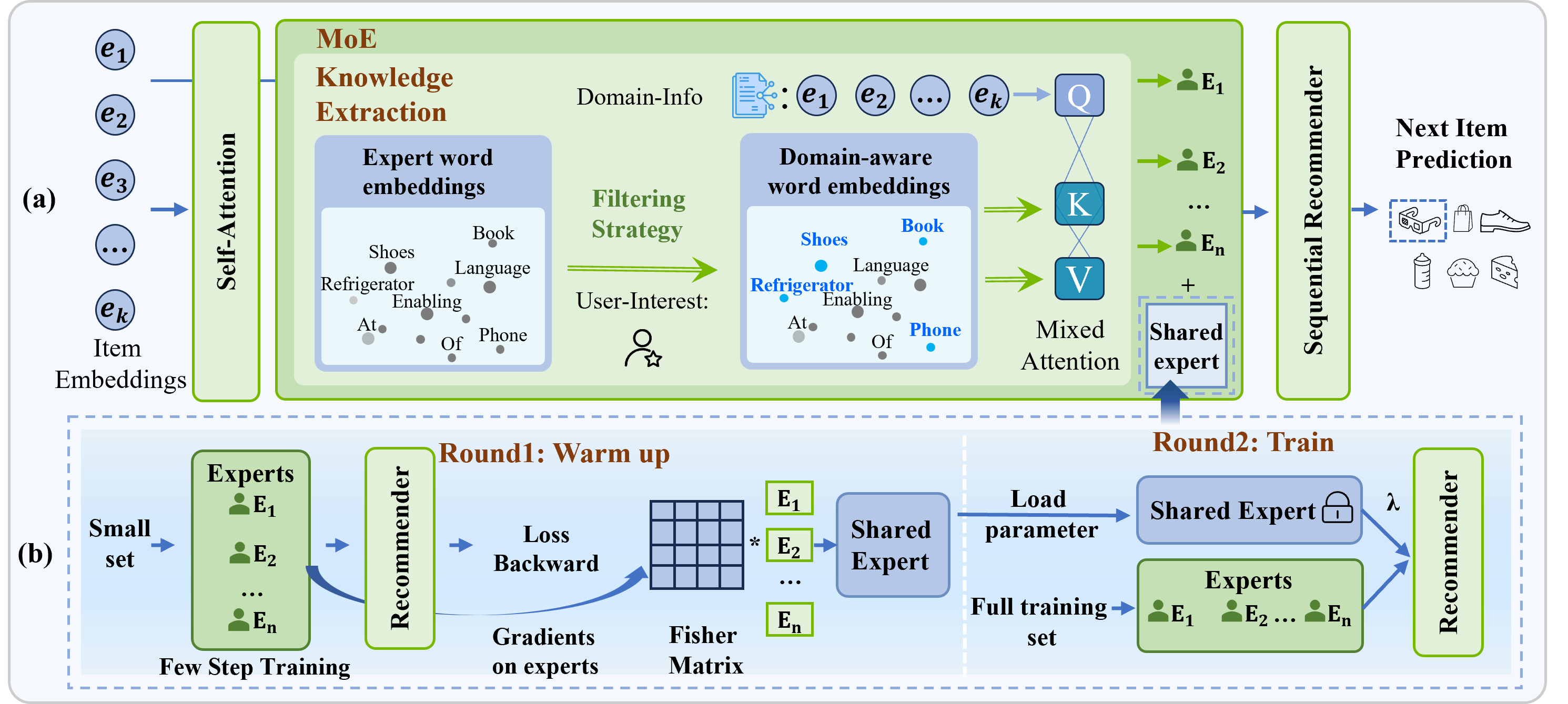}
\captionsetup{width=0.9\textwidth}
\caption{Architecture of MLTFR. (a) Overall framework: each MoE expert leverages a filtered LLM token set, and the aggregated MoE representation is fed into the sequential recommender for next-item prediction. (b) shared expert training: Round 1 estimates Fisher information and builds a semantic consensus matrix (shared expert) via importance-weighted merging. Round 2 loads and freezes the semantic consensus matrix  parameters, then integrates it as an auxiliary branch while optimizing the remaining experts.
} \label{fig2}
\vspace{-10pt}
\end{figure*}

Figure \ref{fig2} summarizes MLTFR as a plug-and-play module inserted on top of an arbitrary ID-based sequential recommender. Concretely, we first embed the input sequence into
$\mathbf{E} = \{\mathbf{e}_1, \ldots, \mathbf{e}_k\} \in \mathbb{R}^{k \times d_{\text{emb}}}$. MLTFR then enriches $\mathbf{E}$ with semantics extracted from multiple LLM vocabularies: For each expert $m$, MLTFR (i) filters a task-relevant subset of tokens from the corresponding LLM vocabulary, and (ii) derives a knowledge matrix 
$\mathbf{W}_{\text{expert}_m} \in \mathbb{R}^{k \times d_{\text{emb}}}$ 
from the selected token embeddings via cross-attention with $\mathbf{E}$. A gating network subsequently performs item-level soft routing to fuse the expert knowledge matrices $\mathbf{W}_{\text{expert}_m} \in \mathbb{R}^{k \times d_{\text{emb}}}$ for expert $m$ into a base semantic representation $\mathbf{W}_{\text{base}}$, which is further combined with a semantic consensus expert to produce the final semantic augmentation $\mathbf{W}_{\text{output}}$. This augmentation is added to the backbone sequence representation and used for next-item prediction.

The rest of this section follows the zoom-in structure of Figure \ref{fig2}: Section \ref{Reprogram_Module} details the per-expert knowledge extraction module (Fig. \ref{fig2}(a)), including  User Information Strategy and the semantic integration layer; Section \ref{Gated Expert Aggregation} describes multi-expert integration via MoE routing (Fig. \ref{fig2}(a));  Section \ref{Fisher} motivates the semantic consensus expert and details its Fisher-based two-round training and frozen fusion. Section \ref{Model Training} presents the training objective (Fig. \ref{fig2}(b)). Section \ref{Large language Model Selection} discusses practical principles for selecting LLMs/tokens for experts.

\subsection{Task-Relevant Token Selection}
\label{Reprogram_Module}

We perform task-relevant token selection independently within each expert, as illustrated in Fig.~\ref{fig2}(a). The goal is to extract a compact set of LLM token embeddings that are informative for sequential recommendation and integrate them with user interaction sequences in a lightweight manner.

\subsubsection{Task-Oriented Semantic Filtering}
\label{DD}

This subsection describes how we filter the LLM vocabulary within each expert to identify token embeddings that are useful for sequential recommendation. Directly using the full token embedding matrix $\mathbf{W}_{\text{voc}} \in \mathbb{R}^{V \times d_{\text{llm}}}$ where $V$ is the vocabulary size is impractical for two reasons. First, it incurs substantial memory overhead, which contradicts the goal of a lightweight and efficient framework. Second, a large portion of the vocabulary consists of tokens that are irrelevant to item recommendation (e.g., rare words or general-purpose terms). Including such tokens introduces noisy signals and weakens the effectiveness of semantic integration.

To address these issues, we adopt a \textbf{User Information Strategy} to select task-relevant token embeddings from each LLM. This strategy relies solely on user-side signals inferred from interaction histories to guide the filtering process, without requiring any additional textual attributes of users or items. As a result, the selected token embeddings are better aligned with user interests and the requirements of sequential recommendation.

\textbf{User Information Strategy.}
We design a user-guided strategy to filter LLM token embeddings according to user interests, with the goal of retaining task-relevant semantic signals while suppressing irrelevant vocabulary noise. In sequential recommendation, item embeddings are iteratively updated during training, and the sequence representation 
$\{\mathbf{e}_1, \mathbf{e}_2, \ldots, \mathbf{e}_k\}$ therefore provides a natural summary of a user’s evolving preferences. 

We represent user interest by the mean-pooled sequence embedding:
\begin{equation}
\mathbf{e}^{\text{interest}} = \mathrm{Mean}(\mathbf{e}_{1}, \mathbf{e}_2, \ldots, \mathbf{e}_k),
\end{equation}
where $\mathbf{e}^{\text{interest}} \in \mathbb{R}^{d_{\text{emb}}}$.
This vector serves as a lightweight proxy for the user’s current preference profile.

To select task-relevant tokens, we retrieve the top-$K$ vocabulary embeddings that are most similar to $\mathbf{e}^{\text{interest}}$. Since the LLM token embedding dimension $d_{\text{llm}}$ is typically larger than the recommender embedding dimension $d_{\text{emb}}$, we adopt a \textbf{dual-space} design. Specifically, token selection is performed in a learnable low-dimensional alignment space, while the original high-dimensional token embeddings are preserved for subsequent semantic integration.
Formally, we first project the LLM vocabulary embeddings into the recommendation embedding space:
\begin{equation}
\mathbf{W}_{\text{voc\_align}} = \mathbf{W}_{\text{voc}} \mathbf{W}_{\text{align}} + \mathbf{b}_{\text{align}};
\label{eq:alignment}
\end{equation}
where $\mathbf{W}_{\text{align}} \in \mathbb{R}^{d_{\text{llm}} \times d_{\text{emb}}} and
\quad
\mathbf{b}_{\text{align}} \in \mathbb{R}^{d_{\text{emb}}}$.
We then compute cosine similarities between aligned token embeddings and the user interest vector:
\begin{equation}
\mathbf{s}=\mathrm{cosine}\!\left(\mathbf{W}_{\text{voc\_align}},\,\mathbf{e}^{\text{interest}}\right)\in\mathbb{R}^{V},
\label{eq:sim}
\end{equation}
where $\mathbf{s}=[s_1,\ldots,s_V]^\top$ and $s_v$ denotes the similarity score of token $v\in\mathcal{V}$, with $V=|\mathcal{V}|$ being the vocabulary size. These similarity scores are used as routing logits, and the Top-$K$ selection is made differentiable as described next.

\textbf{Differentiable sparse selection.}
To make Top-$K$ token selection trainable, we combine a \emph{soft} stochastic router with a \emph{hard} Top-$K$ selector.
Given the similarity score $s_v$ for each vocabulary token $v \in \mathcal{V}$, we inject Gumbel noise to obtain a
\emph{soft routing distribution} $\mathbf{p} \in \mathbb{R}^{|\mathcal{V}|}$:
\begin{equation}
\tilde{s}_v=\frac{s_v+g_v}{\tau},\quad g_v\sim \mathrm{Gumbel}(0,1),\qquad
p_v=\frac{\exp(\tilde{s}_v)}{\sum_{v'\in\mathcal{V}}\exp(\tilde{s}_{v'})}.
\label{eq:gumbel_softmax}
\end{equation}
The distribution $\mathbf{p}$ provides differentiable selection weights over the full vocabulary. We consider the temperature $\tau \in [0.3, 1.0]$ as a reasonable range trading off near-discrete routing and gradient smoothness.

For the forward pass, we derive a \emph{hard} Top-$K$ index set
\begin{equation}
\mathcal{I}_{\mathrm{topK}}=\mathrm{TopK}(\mathbf{p},K),
\label{eq:topk_indices}
\end{equation}
and its dense indicator vector $\mathbf{b}\in\{0,1\}^{|\mathcal{V}|}$, where
$b_v = 1$ if and only if $v \in \mathcal{I}_{\mathrm{topK}}$.
To connect hard selection with soft gradients, we apply the straight-through estimator:
\begin{equation}
\mathbf{b}^{\mathrm{st}}
=(\mathbf{b}-\mathbf{p})\mathrm{.detach}()+\mathbf{p}.
\label{eq:st_topk}
\end{equation}
In the forward pass, $\mathbf{b}^{\mathrm{st}}$ follows the hard Top-$K$ selection
specified by $\mathcal{I}_{\mathrm{topK}}$, while in the backward pass, gradients are
back-propagated through the soft distribution $\mathbf{p}$.
Finally, we retrieve the corresponding original token embeddings from the LLM
vocabulary:
\begin{equation}
\mathbf{W}_{\mathrm{domain}}
=\mathbf{W}_{\mathrm{voc}}[\mathcal{I}_{\mathrm{topK}}]
\in \mathbb{R}^{K\times d_{\mathrm{llm}}}.
\label{eq:original_tensor}
\end{equation}

\subsubsection{Semantic Integration Layer}
\label{E}
After selecting task-relevant token embeddings, the next step is to integrate them with sequential representations for downstream recommendation. A direct combination of token embeddings and sequence embeddings is often suboptimal, as token embeddings are known to exhibit distributional issues such as anisotropy~\cite{li2020sentence}, which can cause misalignment when merged with representations learned for sequential modeling.

To address this issue, we introduce a semantic integration layer that transforms token embeddings into sequence-aligned representations before fusion. This layer operates as an intermediate mapping module, enabling effective interaction between LLM-derived token semantics and backbone sequence representations while preserving the original architecture of the sequential recommender.

Importantly, the integration layer is designed to be modular and non-intrusive, allowing semantic augmentation to be added without modifying the backbone model. This design underpins the \textbf{plug-and-play} nature of MLTFR and enables seamless integration with a wide range of sequential recommendation architectures.
The semantic integration layer is implemented as a multi-head \emph{cross-attention} module. 
Unlike self-attention, the query and key--value pairs are drawn from different sources: the queries are derived from the backbone sequence representations, while the keys and values are derived from the selected LLM token embeddings. 
Formally, the multi-head cross-attention is defined as
\begin{equation}
\mathrm{MultiHead}(\mathbf{Q}, \mathbf{K}, \mathbf{V})
=
\mathrm{Concat}(\mathrm{head}_1, \ldots, \mathrm{head}_h)\mathbf{W}_{O},
\label{eq:mhca}
\end{equation}
where each attention head is computed as
\begin{equation}
\mathrm{head}_i
=
\mathrm{Attention}(\mathbf{E}\mathbf{W}_{q},\ \mathbf{W}_{\mathrm{domain}}\mathbf{W}_{k},\ \mathbf{W}_{\mathrm{domain}}\mathbf{W}_{v}),
\label{eq:head}
\end{equation}
and
\begin{equation}
\mathrm{Attention}(\mathbf{Q}, \mathbf{K}, \mathbf{V})
=
\mathrm{softmax}\!\left(\frac{\mathbf{Q}\mathbf{K}^{\top}}{\sqrt{d_k}}\right)\mathbf{V}.
\label{eq:attn}
\end{equation}

$\mathbf{E}=[\mathbf{e}_1, \mathbf{e}_2, \ldots, \mathbf{e}_k] \in \mathbb{R}^{k \times d_{\text{emb}}}$ denotes the backbone sequence representations, 
$\mathbf{W}_{\mathrm{domain}} \in \mathbb{R}^{K \times d_{\text{llm}}}$ denotes the selected LLM token embeddings,
$\mathbf{W}_q \in \mathbb{R}^{d_{\text{emb}} \times d_{\text{emb}}}$,
and $\mathbf{W}_k, \mathbf{W}_v \in \mathbb{R}^{d_{\text{llm}} \times d_{\text{emb}}}$ are learnable projection matrices. Here $k$ denotes the sequence length, while $K$ denotes the number of selected tokens in the filtered LLM vocabulary (i.e., the Top-$K$ token set size). $\mathbf{W}_O$ is the output projection matrix.
The resulting cross-attention output forms the knowledge matrix
$\mathbf{W}_{\text{expert}_m} \in \mathbb{R}^{k \times d_{\text{emb}}}$ for expert $m$.
\subsection{Semantic Grounding for Cross-LLM Alignment}\label{Mixture-of-Experts}
After deriving expert-specific semantic representations from multiple LLMs, we integrate them for sequential recommendation in a principled manner. While a gated Mixture-of-Experts (MoE) enables flexible expert combination, it does not explicitly enforce semantic consistency across experts, especially when they originate from heterogeneous LLMs or domains. This may lead to redundant or conflicting signals. To address this issue, we introduce a shared semantic grounding mechanism that aligns expert representations and produces a more compact and information-dense semantic augmentation.

\subsubsection{Gated Expert Aggregation}\label{Gated Expert Aggregation}
We adopt a parallel MoE architecture~\cite{jordan1994hierarchical, liu2024beyond} to aggregate semantic representations from multiple experts. The parallel design allows expert aggregation to operate alongside the backbone model without introducing strong inter-module dependencies, thereby preserving efficiency and stability.

We consider $n$ experts, where each expert $m \in \{1,\ldots,n\}$ is represented by a knowledge matrix
$\mathbf{W}_{\text{expert}_m} \in \mathbb{R}^{k \times d_{\text{emb}}}$ derived via semantic integration. Given the backbone sequence representation
$\mathbf{E} \in \mathbb{R}^{k \times d_{\text{emb}}}$, we compute item-level gating weights for each expert as
\begin{equation}
G_m(\mathbf{E})
=
\frac{\exp\!\left(\mathbf{W}_m \mathbf{E}^{\top} + \mathbf{b}_m\right)}
{\sum_{j=1}^{n} \exp\!\left(\mathbf{W}_j \mathbf{E}^{\top} + \mathbf{b}_j\right)},
\end{equation}
where $\mathbf{W}_m \in \mathbb{R}^{1 \times d_{\text{emb}}}$ and $\mathbf{b}_m \in \mathbb{R}^{1 \times k}$ are learnable parameters.
The resulting $G_m(\mathbf{E}) \in \mathbb{R}^{1 \times k}$ assigns a soft routing weight to each item position in the sequence.

Using these item-level routing weights, we aggregate expert outputs as
\begin{equation}
\mathbf{W}_{\text{base}}
=
\sum_{m=1}^{n} G_m(\mathbf{E}) \odot \mathbf{W}_{\text{expert}_m},
\end{equation}
where $\odot$ denotes element-wise multiplication with broadcasting over the embedding dimension.
This dense routing scheme linearly combines all experts for each item position, enabling flexible expert utilization without imposing hard capacity constraints.


\subsubsection{Semantic Consensus Expert}\label{Fisher}
We introduce a shared semantic consensus expert to mediate both intra-LLM and inter-LLM interactions. Unlike random initialization, we construct and train this shared expert using Fisher Information Matrix (FIM), which quantifies the sensitivity of the prediction to expert parameters on a small training set. Specifically, our procedure consists of the following steps.

\textbf{Round 1.} We compute the Fisher information score for each expert using a small-scale training set. The Fisher information characterizes the sensitivity of the final prediction with respect to the expert parameters, and is estimated as the expected squared gradient of the log-likelihood, as defined below:
\begin{equation}
\mathrm{F}_{r}
= \mathbb{E}_{x \sim \mathcal{D}_{r},\, y \sim p_{\theta}(y \mid x)}
\left[
\left(
\nabla_{\theta_{r}} \log p_{\theta}(y \mid x)
\right)^2
\right]
\end{equation}
where $\mathrm{F}_r\in\mathbb{R}$ is a scalar Fisher importance weight that summarizes the overall sensitivity of the model prediction to $\theta_r$. We construct the Semantic Consensus
Expert parameters by taking a Fisher-weighted combination of all experts’ parameters. This aggregation preserves the components that are most important for recommendation across experts and forms a semantic consensus shared among them:
\begin{equation}
\boldsymbol{\Theta}_{\mathrm{SC}}
= \frac{\sum_{r=1}^{n} \mathrm{F}_{r}\,\theta_{r}}
{\sum_{r=1}^{n} \mathrm{F}_{r}},
\end{equation}
where $\theta_r$ denotes the trainable parameters of the $r$-th expert, $\boldsymbol{\Theta}_{\mathrm{SC}}$  means the  Semantic Consensus
Expert.

\textbf{Round 2.} To keep the shared expert as a fixed reference during training, we freeze its parameters and combine its output $\mathbf{W}_{\mathrm{SC}}\in \mathbb{R}^{k \times d_{\text{emb}}}$ with the remaining experts. The final output is computed as a weighted sum of the shared expert output and the aggregated outputs of the other experts, where the weight is applied to the shared expert:
\begin{equation}
 \mathbf{W}_{\text{output}}= \alpha*\mathbf{W}_{\mathrm{SC}}+\mathbf{W}_{\text{base}},
\end{equation}
where $\alpha$ controls the contribution of the shared expert.
\subsection{Model Training}
\label{Model Training}

We train MLTFR using the standard objective adopted by attention-based sequential recommendation models such as SASRec and BERT4Rec.
Given a user interaction sequence, the model is optimized to predict the next item via a sampled softmax formulation with negative sampling.
Specifically, we adopt the binary cross-entropy loss:
\begin{equation}
\mathcal{L}_{\mathrm{rec}}
=
- \sum_{u \in \mathcal{U}} \sum_{t=2}^{k+1}
\left[
\log \sigma\!\left(\hat{y}_{i_t}^{u}\right)
+
\log \left(1 - \sigma\!\left(\hat{y}_{j_t}^{u}\right)\right)
\right],
\end{equation}
where $i_t$ denotes the ground-truth item at position $t$ for user $u$, $j_t$ denotes a negatively sampled item, and $\sigma(\cdot)$ is the sigmoid function.
The prediction score is computed as
\begin{equation}
\hat{y}_{i}^{u}
=
(\mathbf{h}^{u})^{\top}\mathbf{e}_{i},
\end{equation}
where $\mathbf{e}_{i}$ is the item embedding and $\mathbf{h}^{u}$ is the user representation produced by the sequential recommender:
\begin{equation}
\mathbf{h}^{u}
=
\mathrm{SequentialRecommender}\!\left(\mathbf{E} + \mathbf{W}_{\mathrm{output}}\right).
\end{equation}

The complete training procedure of MLTFR is summarized in Algorithm~1 (Appendix~\ref{algorithm}).

\vspace{-5pt}

\section{LLM Selection}\label{Large language Model Selection}

We now discuss how LLMs are assigned to experts while ensuring compatibility across heterogeneous models.
The choice of LLMs is flexible: experts may use different LLMs or share the same backbone.
Our selection follows two guiding principles: \emph{diversity} and \emph{relevance}.

\textbf{Principle of Diversity.}
We favor LLMs that differ in pre-training domains and data distributions.
Combining diverse LLMs broadens semantic coverage and improves robustness under fixed parameter budgets, as complementary knowledge from different domains can be jointly leveraged~\cite{lu2024merge}.
In practice, diversity is assessed by examining the pre-training data sources and task focuses of candidate LLMs.

\textbf{Principle of Relevance.}
We also prioritize LLMs that are closely aligned with recommendation tasks.
Models pre-trained specifically for recommender systems, such as OpenP5~\cite{geng2022recommendation}, capture user--item interactions and sequential patterns more effectively, providing a strong inductive bias for downstream recommendation.

Based on these principles, we select LLMs whose parameter scales are moderate to support multi-model collaboration.
We further ensure that their vocabulary sizes are comparable, which helps prevent optimization bias toward specific experts in the MoE framework.

In our implementation, we use multiple P5 models pre-trained on different recommendation datasets, including \texttt{P5\_beauty\_base}, \texttt{P5\_sports\_base}, and \texttt{P5\_toy\_base}.
To ensure fair evaluation, the datasets used for LLM pre-training do not overlap with those used in our experiments.

For each selected LLM, we extract the token embedding matrix from its embedding layer:
\begin{equation}
\mathbf{W}_{\mathrm{voc}} \in \mathbb{R}^{V \times d_{\text{llm}}},
\end{equation}
which serves as the input vocabulary representation for task-relevant token selection and semantic integration (Section~\ref{Reprogram_Module}),
where $V$ denotes the vocabulary size and $d_{\text{llm}}$ the embedding dimension.
\begin{table}[!t]
\captionsetup{width=\linewidth}
\caption{Details of chosen LLMs (Rec. denotes recommendation).}\label{tab2}
\centering
\resizebox{0.9\linewidth}{!}{
\begin{tabular}{lccc}
\toprule
\multicolumn{1}{c}{Model} & 
\multicolumn{1}{c}{Embedding Size} & 
\multicolumn{1}{c}{Vocab Size} & 
\multicolumn{1}{c}{Scope} \\
\midrule
Llama-2-7b-hf ~\cite{touvron2023llama} & 4,096 & 32,000 & General \\ 
P5 ~\cite{geng2022recommendation} & 768 & 32,100 & Rec. \\
GPT2 ~\cite{radford2019language} & 768 & 50,257 & General \\
BERT ~\cite{kenton2019bert} & 768 & 30,522 & General \\
Qwen3-0.6B~\cite{yang2025Qwen3} &1024&151936&General\\
\bottomrule
\end{tabular}}
\vspace{-15pt}
\end{table}

\section{Experiments}

In this section, we conduct comprehensive experiments to
answer the following research questions:
\begin{itemize}[leftmargin=*]
\item \textbf{RQ1.} How does the performance of MLTFR compare to current state-of-the-art ID-based sequential recommendation models?
\item \textbf{RQ2.} How do different components of the multi-LLM integration affect the performance of MLTFR?
\item \textbf{RQ3.} How sensitive is MLTFR to the weighting of different training objectives?

\item \textbf{RQ4.} How does the performance of MLTFR compare to LLMs that use text templates for sequential recommendation?

\end{itemize}

\subsection{Experiment Settings}


\subsubsection{Training Details}
All models are implemented in PyTorch and optimized using Adam.
The batch size is set to 128, the hidden dimension is 64, and the learning rate is 0.001.
Unless otherwise specified, the number of experts is set to 4.
We apply a dropout rate of 0.2, and the reprogramming layer uses 4 attention heads.
The weighting coefficient $\alpha$ is set to 0.2 by default.
In Round~1, Fisher information scores are computed once training converges. For the shared consensus expert construction, we only fuse \emph{homogeneous} experts instantiated from identical architecture and hidden size to ensure parameter-wise compatibility during fusion.

Experiments are conducted on multiple public datasets from different domains.
We adopt HR@20 and NDCG@20 as evaluation metrics.
Detailed dataset statistics and preprocessing procedures are provided in Appendix~\ref{app:dataset}.
\subsubsection{Baselines}
Our framework is built upon ID-based sequential recommendation and does not rely on auxiliary information such as textual data. Therefore, we selected a series of advanced ID-based sequential recommendation models as comparison targets: SASRec~\cite{kang2018self}, BERT4Rec~\cite{sun2019bert4rec}, LinRec~\cite{liu2023linrec}, CL4SRec~\cite{xie2022contrastive}, FMLPRec~\cite{zhou2022filter}, and BSARec~\cite{shin2024attentive}. The details of baselines are introduced in Appendix~\ref{Baselines}.
\begin{table*}[!ht]
\captionsetup{width=1\linewidth}
\caption{The performance of our framework MLTFR. W/O refers to the situation without using MLTFR. Imp. refers to the overall percentage improvement computed as the geometric mean of HR and NDCG ratios over the baseline.}
\label{tab3}
\centering
\resizebox{0.8\linewidth}{!}{
\begin{tabular}{c| c |ccc ccc ccc}
\toprule
\multirow{2}{*}{\textbf{Dataset}} &\multirow{2}{*}{\textbf{Method}} & \multicolumn{3}{c}{\textbf{SASRec}} & \multicolumn{3}{c}{\textbf{BERT4Rec}} & \multicolumn{3}{c}{\textbf{FMLPRec}} \\
\cmidrule(lr){3-11}
 & & HR & NDCG& Imp. & HR & NDCG& Imp. & HR & NDCG & Imp.  \\
\midrule
\multirow{2}{*}{Office} 
& W/O       & 0.02636 & 0.01076  &--& 0.02413 & 0.00984 &--& 0.02108 & 0.00906 & --\\
\cmidrule(lr){2-11}
& $MLTFR$  & 0.02684  & 0.01104 & 2.21 & 0.02759 & 0.01144 & 15.30 & 0.02275 & 0.00907 & 3.94 \\
\midrule

\multirow{2}{*}{Music} 
& W/O       & 0.04643 & 0.01801 &--& 0.04048 & 0.01568 &--& 0.04348 & 0.01683 &--\\
\cmidrule(lr){2-11}
& $MLTFR$   & 0.04669 & 0.01895 & 2.86 & 0.05027 & 0.01979 & 25.19 & 0.04402 & 0.01723 & 1.81 \\
\midrule

\multirow{2}{*}{Pantry} 
& W/O       & 0.03861 & 0.01490 &--& 0.03222 & 0.01188 &--& 0.03664 & 0.01387 & -- \\
\cmidrule(lr){2-11}
& $MLTFR$  & 0.04366 & 0.01626 & 11.09 & 0.04549 & 0.01770 & 45.04 & 0.04037 & 0.01567 & 11.57 \\
\midrule

\multirow{2}{*}{Beer}
& W/O       & 0.10045 & 0.05115 & -- & 0.11302 & 0.05073 & -- & 0.08560 & 0.04266 & -- \\
\cmidrule(lr){2-11}
& $MLTFR$  & 0.10139 & 0.05122 & 0.54 & 0.10834& 0.04651 & -6.25 & 0.08505& 0.04257 & -0.43 \\
\midrule

\multirow{2}{*}{\textbf{Dataset}} &\multirow{2}{*}{\textbf{MLTFR}} & \multicolumn{3}{c}{\textbf{LinRec}} & \multicolumn{3}{c}{\textbf{CL4SRec}} & \multicolumn{3}{c}{\textbf{BSARec}}  \\
\cmidrule(lr){3-11} 
&  & HR & NDCG& Imp. & HR & NDCG & Imp. & HR & NDCG & Imp. \\
\midrule

\multirow{2}{*}{Office} 
& W/O       & 0.02567 & 0.01026 &--& 0.02535 & 0.01002 & --& 0.02477 & 0.00993 & -- \\
\cmidrule(lr){2-11}
& $MLTFR$  & 0.02546 & 0.01028 & -0.31 & 0.02650 & 0.01021 & 3.21 & 0.02524 & 0.00996 & 1.10 \\
\midrule

\multirow{2}{*}{Music} 
& W/O       & 0.04613 & 0.01715 &--& 0.04674 & 0.01857 &--& 0.04441 & 0.01745 &--\\
\cmidrule(lr){2-11}
& $MLTFR$   & 0.04669 & 0.01720 & 0.75 & 0.04897 & 0.01943 & 4.70 & 0.04515 & 0.01779 & 1.81 \\
\midrule

\multirow{2}{*}{Pantry} 
& W/O       & 0.04324 & 0.01699 &--& 0.03903 & 0.01632 &--& 0.04086 & 0.01494 & --\\
\cmidrule(lr){2-11}
& $MLTFR$   & 0.04450 & 0.01729 & 2.34 & 0.04043 & 0.01631 & 1.75 & 0.04079 & 0.01630 & 4.36 \\
\midrule

\multirow{2}{*}{Beer}
& W/O       &0.05764 & 0.02223 & -- & 0.09371 & 0.04795& -- & 0.09777 & 0.05050 & -- \\
\cmidrule(lr){2-11}
& $MLTFR$   & 0.05905 & 0.02287 & 2.66 &0.09580 &0.04807 & 1.24 & 0.10041 & 0.05163 & 1.47 \\
\bottomrule
\end{tabular}}
\end{table*}

\begin{table}[t] 
\centering
\caption{Ablation of MLTFR variants on Pantry and Music.
$Full$: full model (User filter + SC);  $M_{\text{w/o SC}}$: removing the SC expert.}
\label{tab4}
\setlength{\tabcolsep}{4pt}
\renewcommand{\arraystretch}{1.08}

\resizebox{0.9\linewidth}{!}{
\begin{tabular}{c c cc cc}
\toprule
\multirow{2}{*}{\textbf{Dataset}} & \multirow{2}{*}{\textbf{MLTFR}} &
\multicolumn{2}{c}{\textbf{SASRec}} &
\multicolumn{2}{c}{\textbf{BERT4Rec}} \\
\cmidrule(lr){3-6}
& & HR & NDCG & HR & NDCG \\
\midrule
\multirow{2}{*}{Pantry}
& $Full$ &
\textbf{0.04366} & \textbf{0.01626} &
\textbf{0.04549} & \textbf{0.01770} \\
& $M_{\text{w/o}\text{ SC}}$ &
0.04071 & 0.01516 &
0.04394 & 0.01729 \\
\midrule
\multirow{2}{*}{Music}
& $Full$ &
\textbf{0.04669} & \textbf{0.01895} &
0.05027 & 0.01979 \\
&  $M_{\text{w/o}\text{ SC}}$ 
 &
0.04541 & 0.01755 &
\textbf{0.05109} & \textbf{0.02057} \\
\bottomrule
\end{tabular}
}

\vspace{4pt}

\resizebox{0.9\linewidth}{!}{
\begin{tabular}{c c cc cc}
\toprule
\multirow{2}{*}{\textbf{Dataset}} & \multirow{2}{*}{\textbf{MLTFR}} &
\multicolumn{2}{c}{\textbf{FMLPRec}} &
\multicolumn{2}{c}{\textbf{LinRec}} \\
\cmidrule(lr){3-6}
& & HR & NDCG & HR & NDCG \\
\midrule
\multirow{2}{*}{Pantry}
& $Full$ &
\textbf{0.04037} & \textbf{0.01567} &
\textbf{0.04450} & \textbf{0.01729} \\
&$M_{\text{w/o}\text{ SC}}$ &
0.03966 & 0.01460 &
0.04282 & 0.01622 \\
\midrule
\multirow{2}{*}{Music}
& $Full$ &
\textbf{0.04402} & 0.01723 &
\textbf{0.04669} & 0.01720 \\
& $M_{\text{w/o}\text{ SC}}$ &
0.04379 & \textbf{0.01737} &
0.04616 & \textbf{0.01734} \\
\bottomrule
\end{tabular}
}

\vspace{4pt}

\resizebox{0.9\linewidth}{!}{
\begin{tabular}{c c cc cc}
\toprule
\multirow{2}{*}{\textbf{Dataset}} & \multirow{2}{*}{\textbf{MLTFR}} &
\multicolumn{2}{c}{\textbf{CL4SRec}} &
\multicolumn{2}{c}{\textbf{BSARec}} \\
\cmidrule(lr){3-6}
& & HR & NDCG & HR & NDCG \\
\midrule
\multirow{2}{*}{Pantry}
& $Full$ &
0.04043 & \textbf{0.01631} &
\textbf{0.04079} & \textbf{0.01630} \\
& $M_{\text{w/o}\text{ SC}}$ &
\textbf{0.04099} & 0.01604 &
0.03678 & 0.01436 \\
\midrule
\multirow{2}{*}{Music}
& $Full$ &
\textbf{0.04897} & \textbf{0.01943} &
\textbf{0.04515} & 0.01779 \\
& $M_{\text{w/o}\text{ SC}}$ &
0.04858 & 0.01903 &
0.04467 & \textbf{0.01823} \\
\bottomrule
\end{tabular}
}

\end{table}

\subsection{Overall Performance \textnormal{\textbf{(RQ1)}}}
We implement MLTFR on the aforementioned datasets and compare it with six representative baselines.
Table~\ref{tab3} reports HR and NDCG results with and without MLTFR, together with the relative improvement (Imp.).
Overall, MLTFR yields consistent gains across backbones, while the benefit varies by both the underlying model and the target domain.

\begin{itemize}[leftmargin=*]
    \item \textbf{MLTFR consistently improves sequential recommenders, with the largest gains on BERT4Rec.}
    Averaged over all four datasets, the improvement rates are
    \{SASRec: 4.18\%, BERT4Rec: 19.82\%, FMLPRec: 4.22\%, LinRec: 1.36\%, CL4SRec: 2.73\%, BSARec: 2.44\%\}.
    In particular, BERT4Rec achieves the highest average gain (19.82\%), which is plausible because its bidirectional masking-based training encourages global context aggregation, matching the global receptive-field behavior introduced by the reprogramming layer.
   \item \textbf{The effectiveness is domain-dependent: Pantry benefits the most, while Beer is the weakest.}
Averaged over the six backbones, Pantry achieves the highest improvement (12.69\%).
In contrast, Beer exhibits the smallest overall gain, suggesting a larger domain gap between the semantic token sources and the downstream data.
One plausible reason is that P5 is pre-trained on Amazon-domain recommendation corpora, which are closer to Office/Music/Pantry than Beer, thereby providing more transferable semantic signals.
\item \textbf{HR and NDCG are largely direction-consistent, indicating that the gains are not driven by metric tension.}
Across 24 backbone--dataset settings, HR and NDCG move in the same direction in 87.5\% cases (21/24), including 19 cases where both metrics improve.
Only three settings exhibit opposite trends (one metric slightly increases while the other slightly decreases), and the degradations where both HR and NDCG drop are concentrated on Beer (BERT4Rec and FMLPRec), suggesting that the remaining negative cases are mainly due to domain mismatch rather than an inherent HR--NDCG tension.
\end{itemize}

\begin{figure}[!h]
    \centering
\includegraphics[width=0.45\textwidth, , clip]{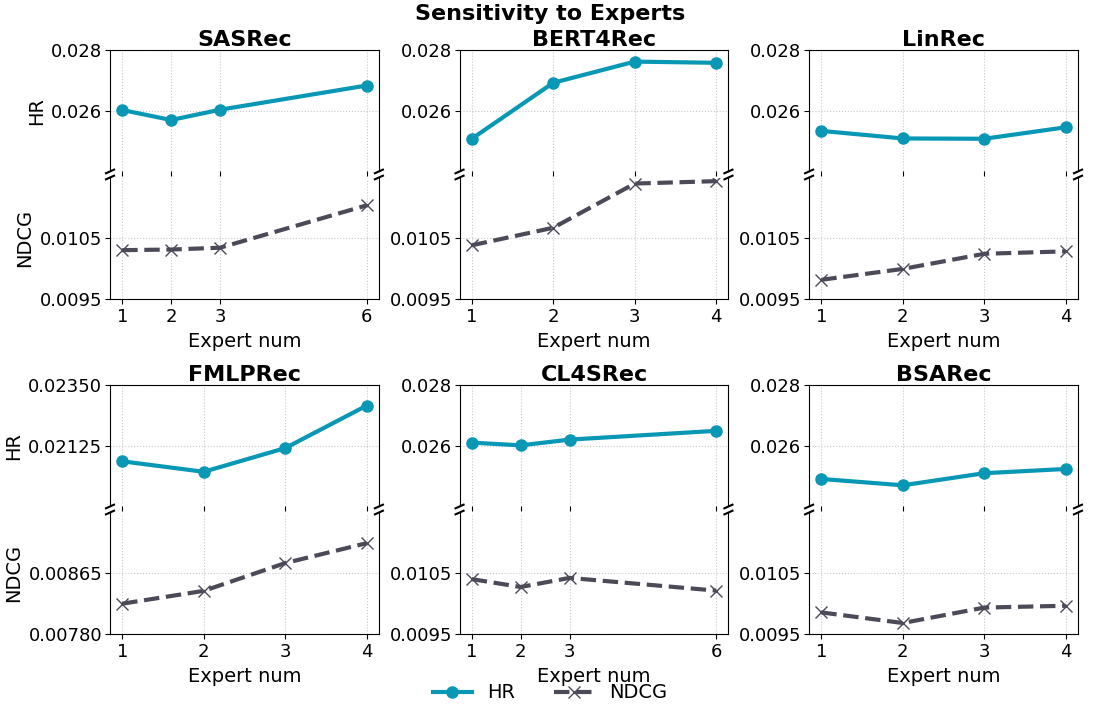}
    \caption{Performance of MLTFR on the Office dataset with six baselines under varying numbers of experts.}
    \label{fig6}
    \vspace{-10pt}
\end{figure}

\subsection{Ablation Study \textnormal{\textbf{(RQ2)}}}
Results in Fig.~\ref{fig:tsc_intro} suggest that leveraging LLM token embeddings provides useful semantic signals. We further ablate the semantic consensus (SC) component in the MoE module to quantify its role in multi-LLM integration. Specifically, we compare the full variant (\textbf{Full}: User filter + Token + SC) with a counterpart that removes the SC expert (\textbf{$M_{\text{w/o SC}}$}) while keeping all other components unchanged. Table~\ref{tab4} reports results on Pantry and Music across six backbone models.

Overall, \textbf{Full} achieves consistent or better performance in most settings, with the most notable gains on Pantry, indicating that SC helps stabilize and consolidate heterogeneous signals from multiple LLM experts. The improvements are more pronounced on backbones with stronger semantic modeling capacity (e.g., BERT4Rec and CL4SRec), suggesting that explicit consensus is particularly beneficial when expert representations are diverse.

In a few cases (mainly on Music), $M_{\text{w/o SC}}$ slightly outperforms \textbf{Full} on individual metrics, implying that enforcing consensus can introduce mild regularization. Nevertheless, the differences are small, and the overall trend supports SC as an effective mechanism for reducing redundancy and mitigating conflicts among multiple LLM experts.

\vspace{-5pt}
\begin{table}[!h]
\centering
\caption{Model performance with Top-1 Accuracy.}
\label{tab:hit_ratios}
\setlength{\tabcolsep}{4pt}
\resizebox{0.9\linewidth}{!}{
\begin{tabular}{@{}lccccc@{}}
\toprule
      & \textbf{BERT4Rec} & \textbf{FMLPRec} & \textbf{LinRec} & \textbf{SASRec} & \textbf{glm4-plus} \\ \midrule
$Pantry$ & 0.27460 & 0.26632 & 0.27095 & 0.27194 & 0.19040 \\
$Office$ & 0.24520 & 0.24360 & 0.23720 & 0.24580 & 0.12140 \\
\bottomrule
\end{tabular}}
\vspace{-10pt}
\end{table}

\subsection{Parameter Analysis \textnormal{\textbf{(RQ3)}}}
We investigate the sensitivity of MLTFR to the number of experts in the MoE module, as well as to the composition of the underlying LLM experts.
Figure~\ref{fig6} reports results on the Office dataset across six backbone sequential recommenders under varying numbers of experts. For a controlled comparison, we build experts by sequentially adding LLMs from an ordered list:
$[\text{Llama},\ \text{BERT},\ \text{P5\_sports\_base},\ \text{P5\_beauty\_base},\ \text{gpt2},\ \text{Qwen3-0.6B}]$,
and use the first $m$ models to form an $m$-expert MoE, where $m \in [1,6]$.

Overall, increasing the number of experts leads to a consistent improvement trend in both HR and NDCG across most backbones.
Nevertheless, the gains are not strictly monotonic: several configurations exhibit performance plateaus or slight degradations when additional experts are introduced.
This behavior indicates that MLTFR benefits not only from a larger expert pool, but also from the semantic complementarity of the added LLMs.
In particular, incorporating experts with redundant or weakly aligned vocabularies may yield diminishing returns, highlighting the importance of expert quality and coverage in multi-LLM integration.

\subsection{Comparison with Black-box LLM Model \textnormal{\textbf{(RQ4)}}}
We examine whether our framework, which does not require item text corpora, can compete with prompt-based LLM recommenders that rely on item titles. For this comparison, we use the latest open-source API of ChatGLM~\cite{glm2024chatglm}, glm4-plus, a bilingual model supporting both Chinese and English. To apply glm4-plus for recommendation, we format the user’s interaction history——candidate items into a prompt, where each item is represented by its title. For fairness, we restrict the candidate pool to the same limited range for both our framework and glm4-plus, and treat glm4-plus as a black-box model without fine-tuning. The top-1 accuracy on Pantry and Office is reported in Table~\ref{tab:hit_ratios}.

As shown in the table, ID-based sequential recommenders equipped with MLTFR substantially outperform the black-box LLM baseline despite not using any textual inputs.
On Pantry, the best-performing backbone improves Top-1 accuracy from 0.190 to 0.275, corresponding to a relative gain of over 44\%.
On Office, the improvement is even more pronounced, with Top-1 accuracy increasing from 0.121 to 0.246, yielding a relative gain exceeding 100\%.
In contrast, although equipped with carefully designed prompts, the LLM occasionally generates responses that do not correspond to valid items in the candidate set, which partially explains its inferior performance under constrained recommendation settings.

\section{Related Work}
\subsection{Sequential Recommendation}

Sequential recommendation infers user preferences based on the items they consume in a time-ordered sequence and predicts the next item they are likely to interact with. The early sequential recommendation models used the Chains (MCs) model~\cite{rendle2010factorizing}, which could only predict the next moment based on the previous moment. Later, deep learning made significant progress in sequential recommendation, with convolutional neural networks (CNN)~\cite{tang2018personalized} and recurrent neural networks (RNN)~\cite{hidasi2015session} successfully capturing local dependencies in sequential data for modeling. However, these models struggled to learn long-range sequential information.

Transformer-based models such as SASRec~\cite{kang2018self} and BERT4Rec~\cite{sun2019bert4rec} addressed the challenge of capturing long-range dependencies and sequence representation through self-attention mechanisms. Specifically, SASRec divides subsequences using a sliding window approach to make full use of the dataset information, while BERT4Rec uses mask to guide the model to learn from contextual information. To better adapt the transformer architecture for sequential recommendation, many variants of attention have emerged. For example, STRec~\cite{li2023strec} replaces self-attention with cross-attention to reduce the model's focus on redundant items, and PATT~\cite{liu2024probabilistic} integrates determinantal point processes (DDPs) into the attention mechanism. In addition, other techniques have impacted the field of sequential recommendation, such as FMLPRec~\cite{zhou2022filter} and DLFS-Rec~\cite{liu2023distribution}, which use frequency domain filtering, and DiffuRec~\cite{li2023diffurec} and DreamRec~\cite{yang2024generate}, which adopt diffusion mechanisms.

Our framework is grounded in sequential recommendation, and MLTFR is compatible with most existing methods in this domain.
\subsection{LLM for Sequential Recommendation}
The temporal nature of sequential recommendation data makes it convenient to integrate with LLMs. Some models use LLMs as the main recommender and design methods to adapt LLMs for recommendation tasks. For example, POD~\cite{li2023prompt} distills discrete templates into continuous vectors, bridging the gap between IDs and words. LLaRA~\cite{liao2023llara} divides tasks into easy and hard parts, gradually unlocking LLMs parameters to address the challenge of LLMs struggling to understand recommendation embeddings. BinLLM~\cite{zhang2024text} uses binary encoding as a means to link text and IDs. LLM-TRSR~\cite{zheng2024harnessing} compresses and summarizes text with LLMs before entering the recommendation process, making it suitable for rich text scenarios.

Semantic-ID approaches such as TIGER~\cite{rajput2023recommender}, IDGenRec~\cite{tan2024idgenrec}, and ActionPiece~\cite{hou2025actionpiece} replace raw item IDs with sequences of discrete semantic tokens from a shared vocabulary, enabling generative recommenders to operate over tokenized item representations instead of large item embedding tables.

Another category of sequential recommendation uses LLMs as a tool to extract textual information and assist in recommendation. SAID~\cite{hu2024enhancing} pre-trains item ID embeddings using the text information encoded by LLMs, aligning information from both domains to enhance recommendation performance. SeRALM~\cite{ren2024enhancing} exploits the text processed by LLMs to enhance ID-based recommendations.
Rather than fine-tuning the entire LLMs as in previous methods, our approach uses only partial LLMs parameters and is inspired by the idea of multi-LLM routing.

\section{Conclusion}
We propose a corpus-free paradigm, Interaction-Guided LLM Knowledge Integration, to eliminate reliance on text templates in LLM-enhanced sequential recommendation. This paradigm introduces a unified framework that seamlessly plugs into foundational sequential recommendation models. Within the framework, we design a user interest-guided token filtering strategy during knowledge extraction, and use reprogramming to implicitly align LLM token embeddings with recommendation semantics. To integrate vocabulary embeddings from multiple LLMs, we employ a MoE architecture, where each expert corresponds to the filtered knowledge extracted from one LLM. On top of MoE, we further introduce a semantic consensus matrix to reconcile redundancy across experts and provide a shared semantic grounding. The resulting module functions as a plug-and-play layer over base sequential recommenders and requires no auxiliary textual attributes from the dataset. Experiments on various sequential recommendation backbones demonstrate the effectiveness of our approach, and ablation studies verify the practical role of LLM vocabularies in improving recommendation performance.





\bibliographystyle{ACM-Reference-Format}
\bibliography{file}
\appendix

\section{Complex Analysis}\label{Complex Analysis}
Our MLTFR incorporates an MoE module on backbone models such as SASRec and BERT4Rec, so we focus on the complexity analysis of the MoE component to demonstrate that the additional time and memory overhead of our paradigm remains within the same order of magnitude as traditional sequential recommendation models.

Let \(N\) be the number of experts, \(T\) be the number of items in each sequence, \(L\) be the number of attention layers in the recommendation backbone model, \(K\cdot d_{\mathrm{llm}}\) be the dimension of the vocabulary matrix, and \(T\cdot d_{\mathrm{emb}}\) be the dimension of the sequence embedding matrix.
The time complexity of the MoE module is
\[
O\!\Big(N\big(T\cdot d_{\mathrm{emb}}^{2} \;+\; K\cdot d_{\mathrm{llm}}\cdot d_{\mathrm{emb}} \;+\; T\cdot K\cdot d_{\mathrm{emb}}\big)\Big),
\]
where \(T\cdot d_{\mathrm{emb}}^{2}\) is the cost of mapping the query matrix (sequence embedding), \(K\cdot d_{\mathrm{llm}}\cdot d_{\mathrm{emb}}\) is the cost of mapping the key and value (vocabulary matrix), and \(T\cdot K\cdot d_{\mathrm{emb}}\) is the cost of self-attention computation.
In contrast, the time complexity of the traditional attention mechanism is
\[
O\!\big(T\cdot d_{\mathrm{emb}}^{2} \;+\; T^{2}\cdot d_{\mathrm{emb}}\big).
\]
The overall time complexity of MLTFR is
\[
O\!\Big(N\big(T\cdot d_{\mathrm{emb}}^{2} \;+\; K\cdot d_{\mathrm{llm}}\cdot d \;+\; T\cdot K\cdot d_{\mathrm{emb}}\big)\Big)
\;+\;
O\!\big(L(T\cdot d_{\mathrm{emb}}^{2} \;+\; T^{2}\cdot d_{\mathrm{emb}})\big).
\]
The additional overhead mainly comes from the LLM's embedding dimension and vocabulary dimension.

In previous approaches, it was necessary to encode the text.
Assuming the average text length of an item is \(A\), the combined prompt length input into the LLMs is \(A\cdot T\), and \(L_{\mathrm{llm}}\) is the number of LLM layers, we estimate the time complexity to process the sequence through the LLMs as
\[
O\!\big(L_{\mathrm{llm}}((AT)^{2}\cdot d_{\mathrm{llm}})\big).
\]
Even without further processing by smaller models, this time overhead is significantly higher than that of MLTFR.
This demonstrates that our paradigm is more suited to fast iteration in industrial application scenarios.

\section{Algorithm of MLTFR}\label{algorithm}
\begin{algorithm}[ht]
\SetAlgoNlRelativeSize{-1}
\caption{MLTFR}
\KwIn{$S$: user sequence, $L$: LLM vocab list, $R$: number of epochs (Round 2), $N$: number of experts, $R_1$: number of epochs (Round 1), $\alpha$: shared expert weight}
\KwOut{$Y$: predicted item}

\textbf{Initialize} sequential recommender $S(p|\theta_S)$, experts $\{E_m(p|\theta_{E_m})\}_{m=1}^{N}$, user filtering strategy $f$, and gating network $G(p|\theta_G)$\;

\BlankLine
\textbf{Round 1: Train experts and build shared expert (SC)}\;
\For{$r = 0$ \KwTo $R_1-1$}{
    \For{$m = 1$ \KwTo $N$}{
        Compute $\mathbf{W}_{expert_m}$ using $L_m$ and $f$\;
    }
    Fuse experts with $G(p|\theta_G)$ to obtain $\mathbf{W}_{base}$ using (11)--(12)\;
    Compute outputs using $S(p|\theta_S)$ and compute $\mathcal{L}_{Rec}$\;
    Update $\theta_S$, $\{\theta_{E_m}\}_{m=1}^{N}$, and $\theta_G$\;
}
Compute Fisher importance $\{F_m\}_{m=1}^{N}$ using (13)\;
Construct shared expert parameters $\theta_{SC}=\frac{\sum_{m=1}^{N}F_m\,\theta_{E_m}}{\sum_{m=1}^{N}F_m}$ and obtain $\mathbf{W}_{SC}$\;
Freeze $\theta_{SC}$\;

\BlankLine
\textbf{Round 2: Train with shared expert (SC)}\;
\For{$r = 0$ \KwTo $R-1$}{
    \For{$m = 1$ \KwTo $N$}{
        Compute $\mathbf{W}_{expert_m}$ using $L_m$ and $f$\;
    }
    Fuse experts with $G(p|\theta_G)$ to obtain $\mathbf{W}_{base}$ using (11)--(12)\;
    Compute $\mathbf{W}_{output}=\mathbf{W}_{base}+\alpha\,\mathbf{W}_{SC}$ using (15)\;
    Compute outputs using $S(p|\theta_S)$ and compute $\mathcal{L}_{Rec}$\;
    Update $\theta_S$, $\{\theta_{E_m}\}_{m=1}^{N}$, and $\theta_G$ (keep $\theta_{SC}$ frozen)\;
}

\BlankLine
\textbf{Inferencing:}\;
\For{$m = 1$ \KwTo $N$}{
    Compute $\mathbf{W}_{expert_m}$ using $L_m$ and $f$\;
}
Fuse experts with $G(p|\theta_G)$ to obtain $\mathbf{W}_{base}$\;
Compute $\mathbf{W}_{output}=\mathbf{W}_{base}+\alpha\,\mathbf{W}_{SC}$\;
Compute outputs using $S(p|\theta_S)$ and return $Y$\;

\end{algorithm}

\section{Datasets}\label{app:dataset}
\begin{table}[!t]
\captionsetup{width=0.86\linewidth}
\caption{The Statistics of datasets. Av. n means the average length of sequences.}\label{tab1}
\centering
\begin{tabular}{lcccc}
\toprule
Dataset & \#Users & \#Items & Av. n & Density \\
\midrule
Office & 77,284 & 27,467 & 4.34491 & $1.5819\times 10^{-4}$ \\ 
Music & 19,513 & 10,445 & 4.52729 & $4.3344\times 10^{-4}$ \\
Pantry & 7,123 & 4,794 & 4.23838 & $8.8410\times 10^{-4}$ \\
Beer &16,273  & 110,236 & 22.09021 &$2.0039\times 10^{-4}$   \\
\bottomrule
\end{tabular}
\end{table}
Most current sequential recommendation models use datasets from the e-commerce domain. These datasets contain clear interaction timestamps, which can be processed into user sequence patterns to predict future items. We chose the Amazon Review dataset~\cite{ni2019justifying}, which originates from the Amazon e-commerce website and includes multiple sub-datasets from different domains, making it a versatile dataset. We selected the ``Office Products", ``Prime Pantry", and ``Musical Instruments" sub-datasets, and for each dataset, we used a 5-core dense version. Additionally, to demonstrate the cross-platform adaptability of our model, we also chose the ``RateBeer"  dataset ~\cite{rappaz2017bartering}. 

The most recent item interaction of each user is used for evaluation. When splitting the training and test sets, we removed sequences in the training set that became too short (length less than 3) after splitting. The details of the data sets are shown in Table \ref{tab1}.

\section{Details of Baselines}\label{Baselines} 
\begin{itemize}[leftmargin=*]
\item \textbf{SASRec}~\cite{kang2018self} is a personalized recommendation model.
It uses stacked unidirectional attention blocks to predict
items sequentially.
\item \textbf{BERT4Rec}~\cite{sun2019bert4rec} adopts the Cloze objective, predicting a randomly masked item in the sequence through joint analysis of its preceding and succeeding contexts.
\item \textbf{LinRec}~\cite{liu2023linrec} proves transformer-based recommendation models by using an L2-normalized linear attention mechanism instead of traditional dot-product
attention.
In this paper, LinRec is combined with the prediction
mechanism of BERT4Rec, and will be referred to as LinRec
hereafter.
\item \textbf{CL4SRec}~\cite{xie2022contrastive} incorporates contrastive learning into a sequential recommendation, based on SASRec. It samples three methods, masking, reordering, and cropping, to create augmented data samples for contrast.
\item \textbf{FMLPRec}~\cite{zhou2022filter} improves the attention mechanism by replacing multi-head attention with MLP layers in the frequency domain, reducing the risk of overfitting.

\item \textbf{BSARec}~\cite{shin2024attentive} provides a recommendation module that combines self-attention mechanisms with Fourier transforms to alleviate the oversmoothing problem inherent in traditional self-attention. 
\end{itemize}

\end{document}